\documentclass[aps,prb,twocolumn,floatfix,amsmath,amssymb]{revtex4}
\usepackage[final]{graphicx}
\usepackage{bm}
\usepackage{float}
\usepackage{afterpage}

\begin{document}

\title{Interaction effect in two-dimensional Dirac fermions}

\author{Yongfei Jia, Huaiming Guo$^{*}$, Ziyu Chen}

\affiliation{Department of Physics, Key Laboratory of Micro-nano
Measurement-Manipulation and Physics, Beihang University, Beijing, 100191, China}

\author{Shun-Qing Shen}
\affiliation{Department of Physics, The University of Hong Kong, Pokfulam Road, Hong Kong}

\author{Shiping Feng}

\affiliation{Department of Physics, Beijing Normal University, Beijing, 100875, China}

\begin{abstract}
Based on the two-dimensional $\pi-$ flux model, we study the interaction effects both in nontrivial massive and massless Dirac fermions with numerical exact diagonalization method. In the presence of the nearest and next-nearest neighbor interactions: for nontrivial massive Dirac fermion, the topological phase is robust and persists in a finite region of the phase diagram; while for massless Dirac fermion, charge-density-wave and stripe phases are identified and the phase diagram in $(V_1, V_2)$ plane is obtained. When the next-next-nearest neighbor interaction is further included to massless Dirac fermion, the topological phase expected in the mean-field theory is absent. Our results are related to the possibility of dynamically generating topological phase from the electronic correlations.
\end{abstract}


\maketitle

\section{Introduction}
Topological insulator (TI) is a new class of time-reversal invariant quantum phase, which is topologically distinct from the trivial band insulator. The experimental discovery of such phase has simulated the study of topological properties. It has been shown that TI has many exotic physical properties, which make it have potential applications in the fields of spintronics, quantum computing, etc \cite{rev1,rev2,rev3,rev4}.

Recently due to the rich physics and fundamental interest, the correlation effect in the context of TI has been explored \cite{int1,int2}. One aspect of such study is to consider the interplay between the interaction and the intrinsic spin-orbit coupling. Based on the Kane-Mele-Hubbard model \cite{km1, km2}, many numerical and analytical works are carried out \cite{kmu1, kmu3, kmu5,kmu6, kmu7, kmu8, th1,th2,th3,th4}. In these studies consistent results are obtained, such as: the stability of TI to small interaction; the magnetic phase at strong correlation; the existence of quantum spin liquid phase; et al. Another aspect of studying the correlation effect is to dynamically generate spin-orbit coupling from the electronic correlations, resulting topological Mott insulator (TMI) \cite{tmi1}. At the mean-field level, TMI has been found in systems with different geometries \cite{tmi2,tmi3,tmi4,tmi6,tmi7,tmi8,tmi9,tmi10}.
Since the present studies on TMI are limited to mean-field approximation, it is necessary to provide direct evidences from exact numerical methods.

In this paper, based on the two-dimensional $\pi-$ flux model, we study the interaction effects both in nontrivial massive and massless Dirac fermions. With numerical exact diagonalization (ED) method, we calculate the Chern number, the fidelity metric, the energy spectrum, the static structure factor (SSF) and the electrons' distribution to characterize different phases. For nontrivial massive Dirac fermion, the topological phase persists in a finite region of the $(V_1,V_2)$ phase diagram. For massless Dirac fermion and in the presence of nearest neighbor (NN) and next-nearest neighbor (NNN) interactions, charge-density-wave (CDW) and stripe phases are identified and the phase diagram in $(V_1, V_2)$ plane is obtained. However when the next-next-nearest neighbor (NNNN) interaction is further included, the topological phase expected in the mean-field theory isn't identified. Our results are related to the possibility of dynamically generating topological phase from the electronic correlations.

\begin{figure}[htbp]
\centering
\includegraphics[width=7.5cm]{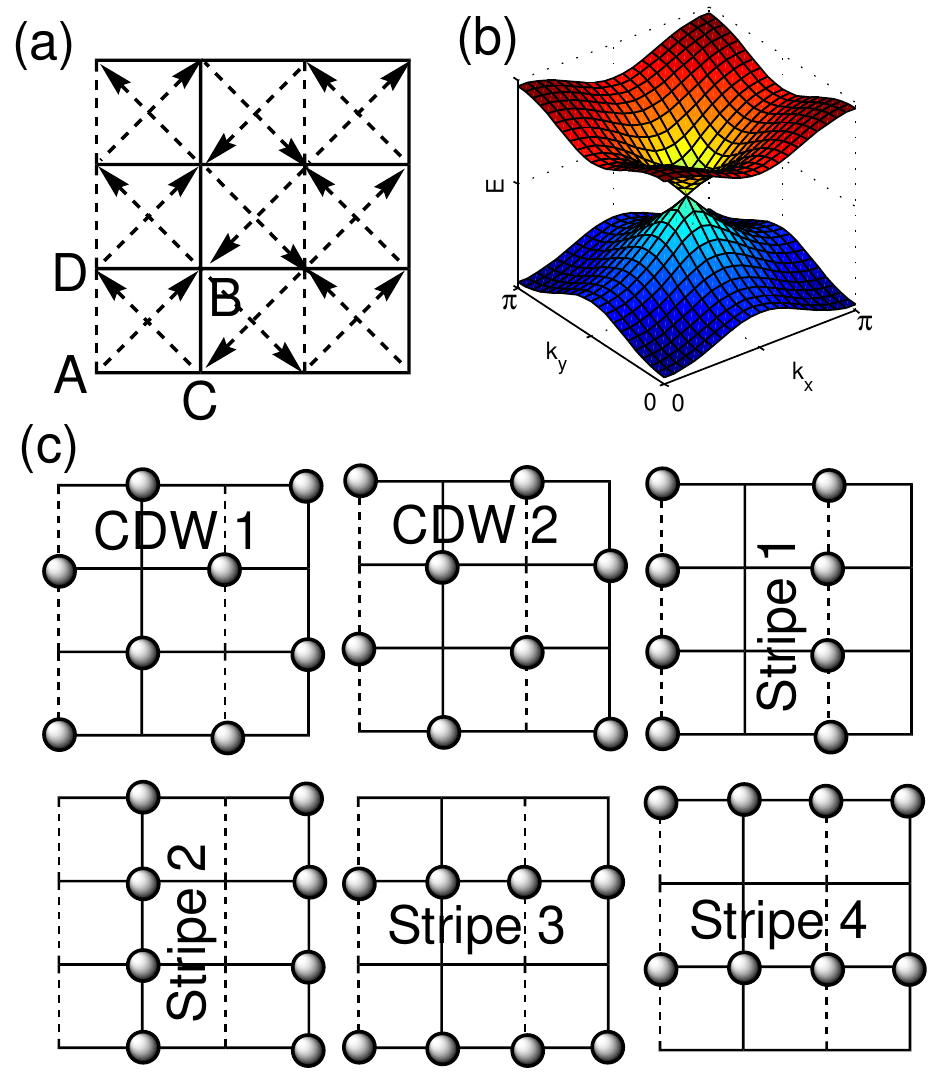}
\caption{(Color online)(a) The hopping amplitudes of Hamiltonian Eq.(\ref{eq1}): nearest neighbor solid links $t_1$, vertical dashed links $-t_1$, diagonal links $it_2$ in the direction of the arrow. (b) The Dirac point in the spectrum when $t_2=0$. (c) the two degenerate configurations with CDW order and the four configurations with the stripe order.}\label{lattice}
\end{figure}
\section{The model and method}
We consider a $\pi-$ flux model on square lattice with a tight-binding Hamiltonian \cite{model1,model2},
\begin{equation}\label{eq1}
H_0= \sum_{ij}t_{ij}e^{i\chi_{ij}}c^\dag_{j}c_{i},
\end{equation}
where $c^\dag_{i}$ and $c_{i}$ are the annihilation and
creation operators at site ${\bf r}_i$.  When the sites $i$ and $j$ are NN neighbors, $t_{ij}=t_1$ and $\chi_{i,i+\hat{x}}=0, \chi_{i,i+\hat{y}}=\pi i_x$. While for the NNN case, $t_{ij}=t_2$ and $\chi_{i,i+\hat{x}+\hat{y}}=\chi_{i+\hat{x},i+\hat{y}}=\pi i_x$. The phase factors are shown in Fig.\ref{lattice}(a), in which a two-site unit cell $(A,C)$ can be chosen.
In the reciprocal space, the Hamiltonian is written as $H_0=\sum_{\bf{k}}\psi_{\bf{k}}^{\dagger}H_0(\bf{k})\psi_{\bf{k}}$ with $\psi_{\bf{k}}=(c_A,c_C)^{T}$ and
\begin{eqnarray*}
H_0({\bf k})=2t_1\cos{k_x}\sigma_{x}-2t_1\cos{k_y}\sigma_{z}-4t_2\sin k_x sin k_y \sigma_{y},
\end{eqnarray*}
with $\sigma_{x,y,z}$ the Pauli matrices. The energy spectrum is given by
\begin{eqnarray*}
  E_{\bf k} &=& \pm \sqrt{4t_1^{2}(\cos^2 k_x+\cos^2 k_y)+16t_2^2\sin^2 k_x \sin^2 k_y},
\end{eqnarray*}
 which is symmetric around zero and has a gap $4|t_2|$ at the two inequivalent Dirac points ${\bf K}=(\pi/2,\pm \pi/2)$. In the vicinity of the two Dirac points, the low-energy Hamiltonian is described by the Dirac equations:
 \begin{eqnarray}\label{low}
 h_{0\alpha}=-2 t_1 k_x\sigma_x +\alpha 2 t_1 k_y \sigma_z-\alpha 4t_2\sigma_y,
 \end{eqnarray}
 with the valley index $\alpha=\pm 1$.
In order to include the staggered CDW and stripe orders, a four-site unit cell $(A,B,C,D)$ can be chosen. Then in the reciprocal space, with the reduced Brillouin zone $\{{\bf k}: |k_x|,|k_y|\leq \pi/2 \}$, the Hamiltonian writes as $H'_0=\sum_{\bf{k}}\psi_{\bf{k}}'^{\dagger} H'_0(\bf{k})\psi_{\bf{k}}'$ with $\psi_{\bf{k}}'=(c_A,c_B,c_C,c_D)^{T}$ and
\begin{eqnarray*}
H'_0({\bf k})&=&2t_1\cos{k_x}\sigma_{x}\otimes I+2t_1\cos{k_y}\sigma_{y}\otimes \sigma_{y} \\
    &&-4t_2\sin k_x sin k_y \sigma_{z}\otimes \sigma_{y}.
\end{eqnarray*}

For $t_2 \neq 0$ the system is topological with gapless states associated with the edges traversing the gap. The topological phase can be characterized by the Chern number, which is defined as \cite{kmu3},
\begin{eqnarray}\label{eq2}
C=\dfrac{1}{2\pi i}\int_{BZ}d^2k F_{12}({\bf k}).
\end{eqnarray}
Here the field strength $F_{12}({\bf k})=\partial_{k_1} A_2({\bf k})-\partial_{k_2} A_1({\bf k})$ with the Berry connection $A_{1(2)}({\bf k})=\langle n({\bf k}) |\partial _{k_{1(2)}}|n({\bf k}) \rangle$ and $|n({\bf k})\rangle$ the normalized wave function. An analytical calculation from Eq.(\ref{low}) yields $\frac{1}{2} \textrm{sgn}(t_2/t_1)$ for each Dirac point, thus the Chern number equals $\textrm{sgn}(t_2/t_1)$ \cite{jiexi}. Using an numerical approach suggested in Ref.(32), the Chern number of the system is calculated and equals $1(-1)$ for $t_2>0 (<0)$, which is consistent with the analytical result. Alternatively since the system possesses inversion symmetry,  the topological property can be determined by the product of the parites of the wave functions at the four time-reversal invariant momenta $\Gamma_i=(n_i \pi/2) \hat{x}+(m_i \pi/2) \hat{y}$ with $n_i,m_i=0,1$ \cite{fu, guo1}. If we select site $A$ of the unit cell as the center of inversion, the parity operator writes as ${\cal P}[\psi_A(\textbf{r}),\psi_B(\textbf{r}),\psi_C(\textbf{r}),\psi_D(\textbf{r})]=[\psi_A(\textbf{-r}),\psi_B(\textbf{-r}-2\hat{x}-2\hat{y}),\psi_C(\textbf{-r}-2\hat{x}),\psi_D(\textbf{-r}-2\hat{y})]$. In the momentum space it becomes ${\cal P}_{\bf k}=\textrm{diag}(1,e^{-2i(k_1+k_2)},e^{-2ik_1},e^{-2ik_2})$. We calculate the product of the parities of the occupied states and it is $-1$ for the topological phase \cite{note1}.

In the following of the paper, we use ED to study the effect of interactions in Eq.(\ref{eq1}). In this method, the wave functions and eigenvalues of several lowest states can be directly obtained. To characterize the topological property of the system, we calculate the Chern number and fidelity metric of the ground state. In an interacting system, the Chern number can be defined using the twisted boundary phase $\theta_1, \theta_2$, i.e., replacing $k_1,k_2$ with $\theta_1, \theta_2$ in Eq.(\ref{eq2}). The fidelity metric $g$ is defined as $g(V,\delta V)=\frac{2}{N}\frac{1-F(V,\delta V)}{(\delta V)^2}$ with the fidelity $F(V,\delta V)=|\langle \Psi_{0}(V)|\Psi_{0}(V+\delta V)$ the overlap of the ground-state wave functions at $V$ and $V+\delta V$ \cite{kmu3}. In the following we set $t_1=1$ as the energy scale and all ED calculations are performed on $4\times 4$ system.
\section{Effect of interaction to nontrivial massive Dirac fermion}

We study the effect of interaction in Hamiltonian Eq.(\ref{eq1}) and add NN and NNN interactions,
\begin{eqnarray*}
  H_{int} &=&\sum_{\langle ij \rangle}V_{1}n_{i}n_{j}+\sum_{\langle \langle ij \rangle \rangle}V_{2}n_{i}n_{j}
\end{eqnarray*}

\begin{figure}[htbp]
\centering
\includegraphics[width=7cm]{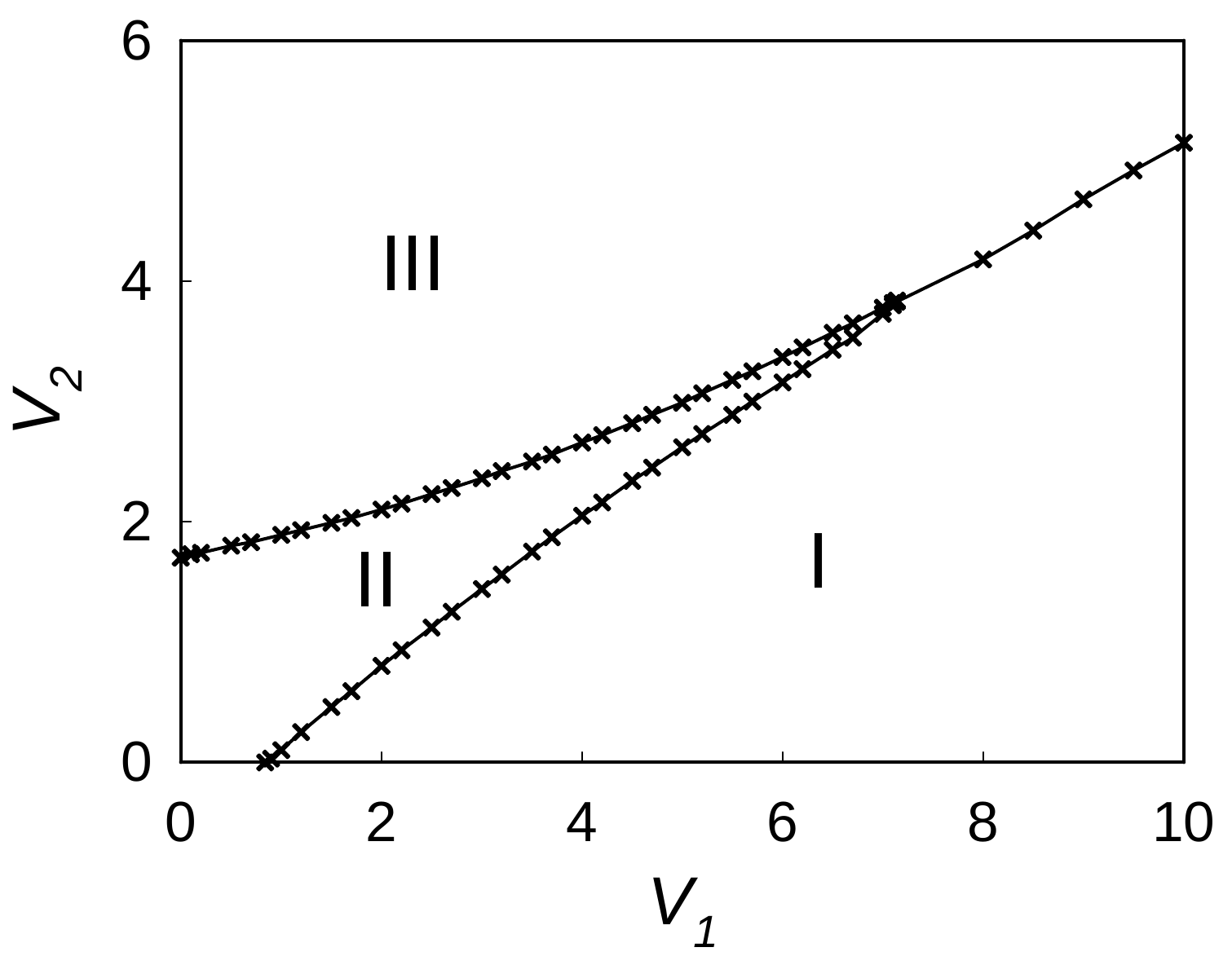}
\caption{(Color online)The phase diagram in $(V_1,V_2)$ plane. Here $t_2=0.1$ at which the non-interacting system is topological. Region I: trivial insulator with CDW order. Region II: topological phase with Chern number $1$. Region III: phase with four-fold degenerate ground state. The size of Region II shrinks as $t_2$ decreases.}\label{phase}
\end{figure}

\begin{figure}[htbp]
\centering
\includegraphics[width=8cm]{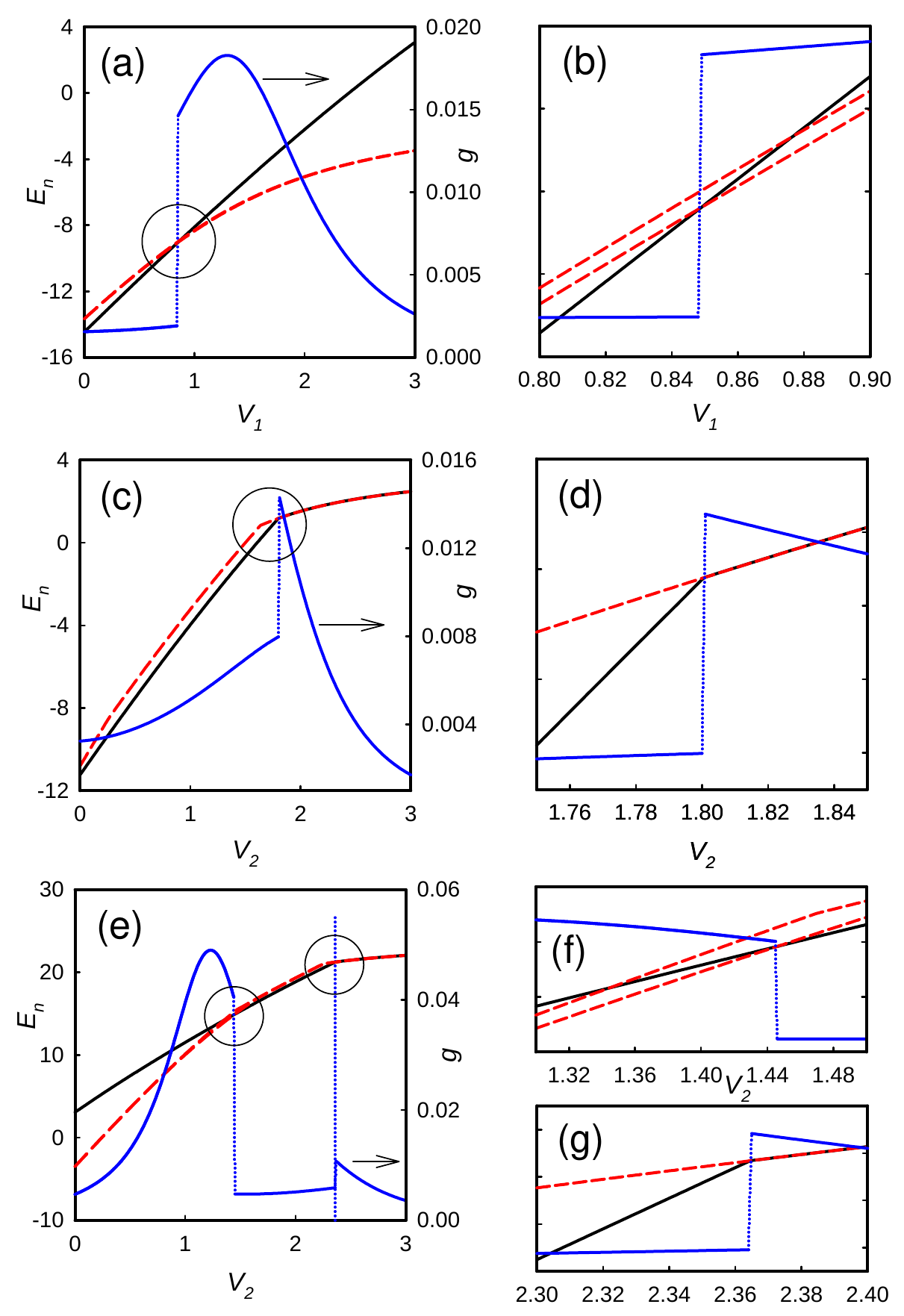}
\caption{(Color online)The eigenenergies of several lowest states and the fidelity metric. (a), (b): $V_2=0$; (c), (d): $V_1=0.5$; (e), (f), (g): $V_1=3$. (b), (d), (f) and (g) are the zooming views near the transition points of the corresponding left figures, which are marked by the circles in (a), (c) and (e). Here $t_2=0.1$. In (c), (d), (e), (f) and (g), at small $V_2$ the red dashed lines correspond to the two states with CDW order (if not split it is two-fold degenerate).}\label{metric}
\end{figure}

By calculating the Chern number of the ground state of the total Hamiltonian $H_0+H_{int}$, the phase diagram is obtained.  As shown in Fig.\ref{phase}, there are three different regions: $\uppercase\expandafter{\romannumeral1}$: the trivial insulator with CDW order; $\uppercase\expandafter{\romannumeral2}$: the topological phase with Chern number $1$; $\uppercase\expandafter{\romannumeral3}$: the phase with four-fold degenerate ground state.

At $V_1=V_2=0$, the system is in a topological phase with quantized Hall conductance $\sigma_{xy}=C e^2/\hbar$ with the Chern number $|C|=1$. As $V_1$ is increased, the system is driven to a trivial insulator with CDW order. The Chern number change from $1$ to $0$ at the critical value $V_1^c=0.84$. In the energy spectrum as shown in Fig.\ref{metric}(a), for $V_1<V_1^c$ the energy of the ground state is gapped from other states. For $V_1>V_1^c$ the gap reopens, but the ground state becomes nearly two-fold degenerate. Further calculations of the distribution of the particles show that they correspond to the two different configurations of staggered CDW order. In Fig.\ref{metric}(b), the three lowest eigenenergies of the system is plotted near the transitional point. The states denoted by the red dashed lines are with CDW order and they are split due to the presence of the hopping terms. The figure clearly show the crossing between the states, which means the disappearance of the topological phase. The obtained critical value is consistent with that from the Chern number. To characterize the topological phase transition, we also calculate the fidelity metric $g$ and the curve shows a jump at the critical interaction.

In Fig.\ref{metric}, we also show the energy spectrum and the fidelity metric $g$ at $V_1=0.5,3$, which are representative cuts in the phase diagram Fig.\ref{phase}. In all cuts the results are consistent with those from calculating the Chern number. At $V_1=0.5$, the system at $V_2=0$ is topological. The NNN interaction drives the topological system to a phase with four-fold degeneracy. At $V_1=3$ the system at $V_2=0$ is a trivial insulator with CDW order. Surprisingly the inclusion of NNN interaction can recover the topological property in a finite region.

To understand the above results, we present a qualitative explanation from a mean-field viewpoint. The interaction can be decoupled in the direct and exchange channel,
\begin{eqnarray*}
n_i n_j\approx &&n_i \langle n_j \rangle +\langle n_i \rangle n_j - \langle n_i \rangle \langle n_j \rangle  \\
&&-\langle c_j^{\dagger} c_i \rangle c_i^{\dagger} c_j -\langle c_i^{\dagger} c_j
\rangle c_j^{\dagger} c_i + \langle c_j^{\dagger} c_i \rangle  \langle c_i^{\dagger} c_j \rangle.
\end{eqnarray*}
If the CDW and stripe orders are considered, we have the ansatz $\langle n_i\rangle=1/2+\rho (-1)^{i_x+i_y}+\nu (-1)^{i_x}$. Let the value of $\langle c_i^{\dagger} c_j\rangle$ is $\delta t$ for NN and $\lambda$ for NNN, then this procedure yields a mean-field Hamiltonian which writes,
\begin{eqnarray*}
 H'_0({\mathbf{k}}) & =&2(t_{1}-\delta t V_{1})\cos k_{x}\sigma_{x}\otimes I \\
 &+&2(t_{1}-\delta t V_{1})\cos k_{y}\sigma_{y}\otimes\sigma_{y} \\
 &-&4(t_{2}-\lambda V_{2})\sin k_{x}\sin k_{y}\sigma_{z}\otimes\sigma_{y}\\
 &-&4\rho(V_{1}-V_{2})\sigma_{z}\otimes I+4\nu V_{2}\sigma_{z}\otimes\sigma_{z}+C_0
\end{eqnarray*}
with the constant $C_0=\frac{N(V_{1}+V_{2})}{2}+2N[(V_{1}-V_{2})\rho^{2}+V_{2}\nu^{2}]+2NV_{1}\delta t^{2}+2NV_{2}\lambda^{2}$. Compared to $H_0({\bf k})$, the hopping amplitudes $t_1$ and $t_2$ are modified and two additional terms related to the CDW and stripe orders appear. Then the phase diagram can be qualitatively explained by including the two additional terms to $H_0({\bf k})$. When only the term $\rho'\sigma_{z}\otimes I$ is included, a topological phase transition is driven by this term, as shown in Fig.4 (a). However when the term $\nu'\sigma_{z}\otimes\sigma_{z}$ is also included, the topological phase can be recovered and further be broken by larger $\nu'$. Since generally the value of $\rho' (\nu')$ increases with $V_1 (V_2)$, the phase diagram is qualitatively explained. Although in a more rigorous way the mean-field parameters should be determined self-consistently, the underlying physics is the same.
\begin{figure}[htbp]
\centering
\includegraphics[width=7cm]{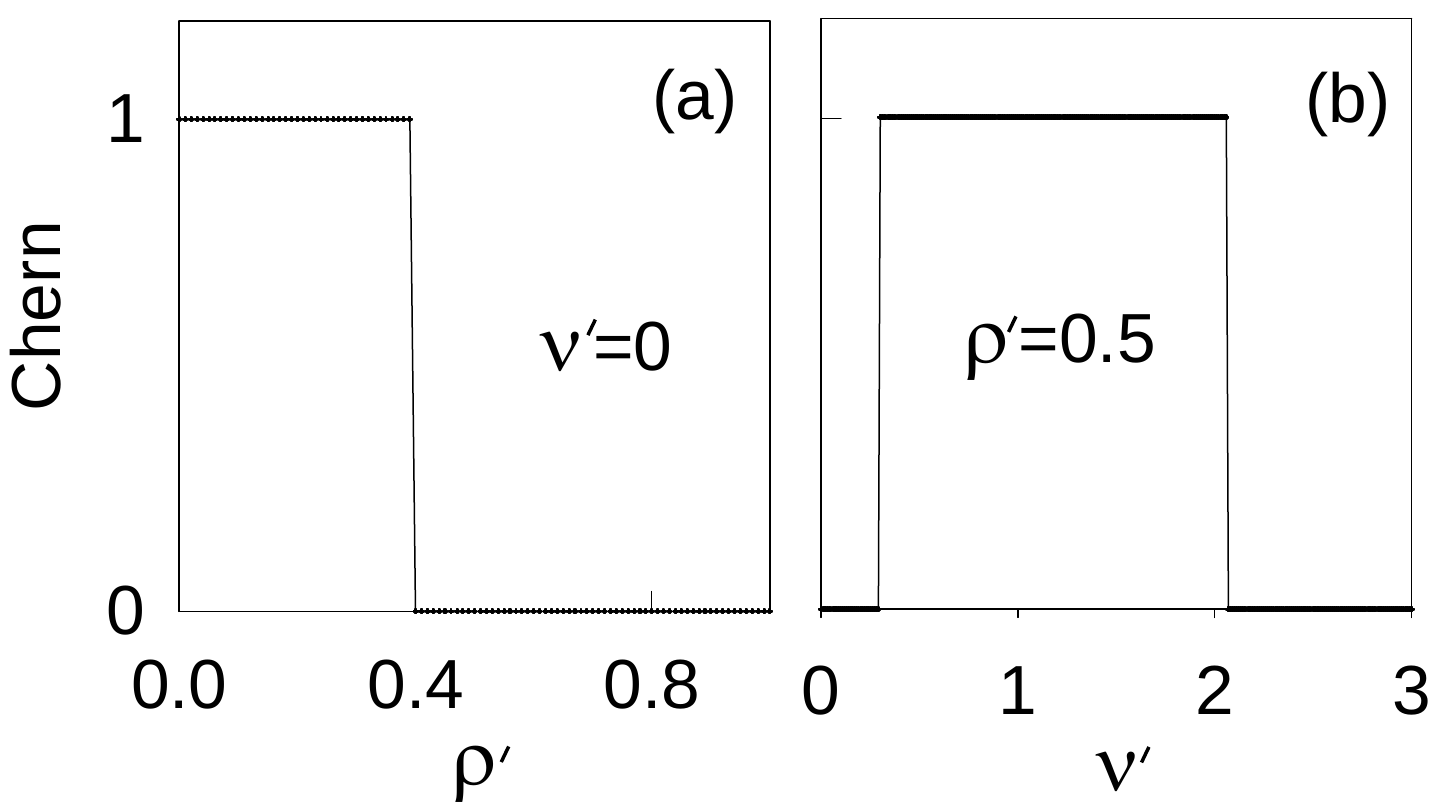}
\caption{(Color online) (a) The Chern number vs $\rho'$ at $\nu'=0$. (b) The Chern number vs $\nu'$ at $\rho'=0.5$.}\label{fig}
\end{figure}

\section{Effect of interaction to massless Dirac fermion}
\begin{figure}[htbp]
\centering
\includegraphics[width=8cm]{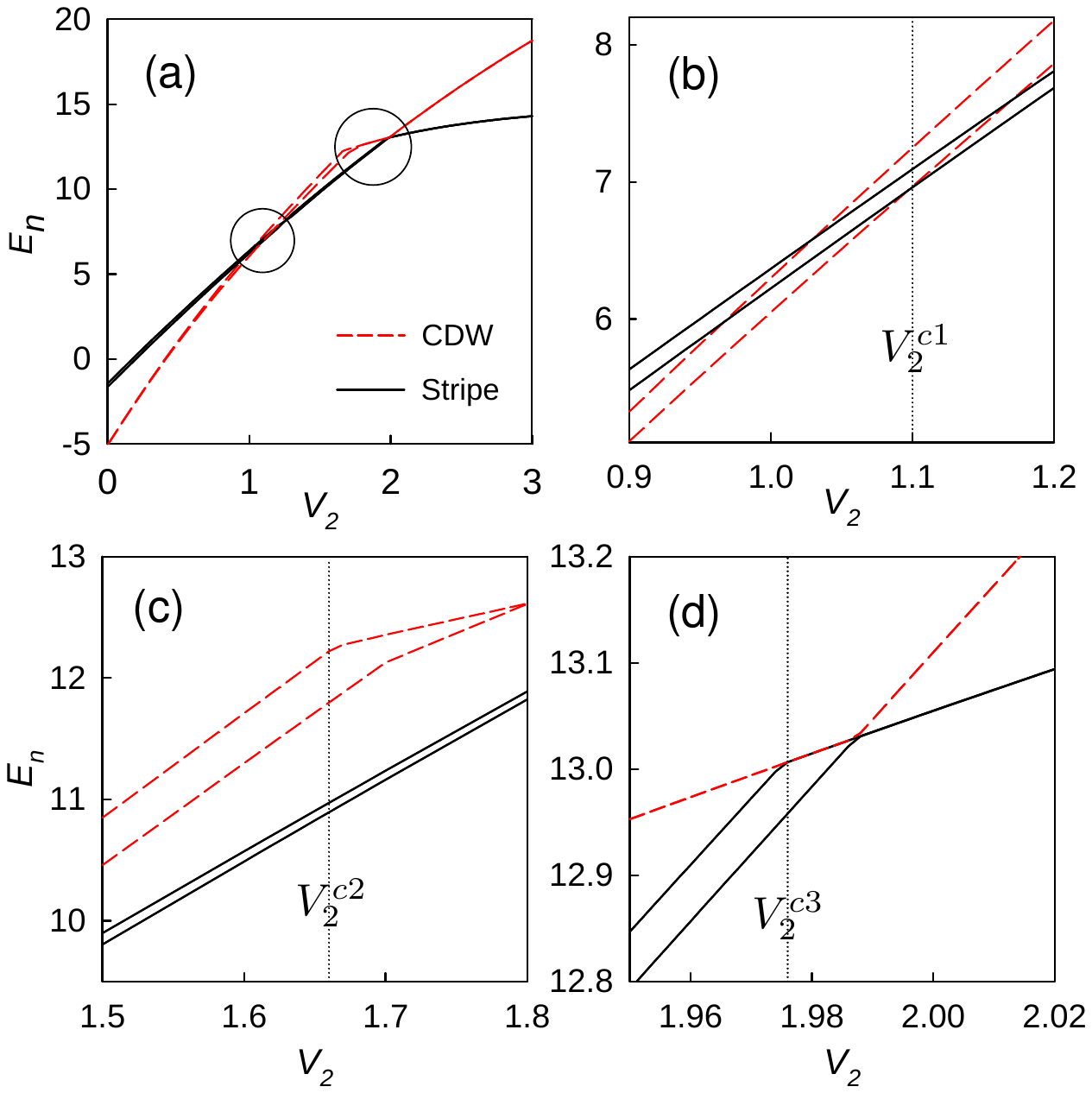}
\caption{(Color online)(a) The eigenenergies of the six lowest states of Eq.(\ref{eq3}) at $V_1=2$. (b), (c) and (d): the zooming views near the transition points, which are marked by the circles in (a). The red dashed lines are for the two CDW states and the black solid lines are for the four stripe states (if splitted each is two-fold degenerate).}\label{en}
\end{figure}

\begin{figure}[htbp]
\centering
\includegraphics[width=7cm]{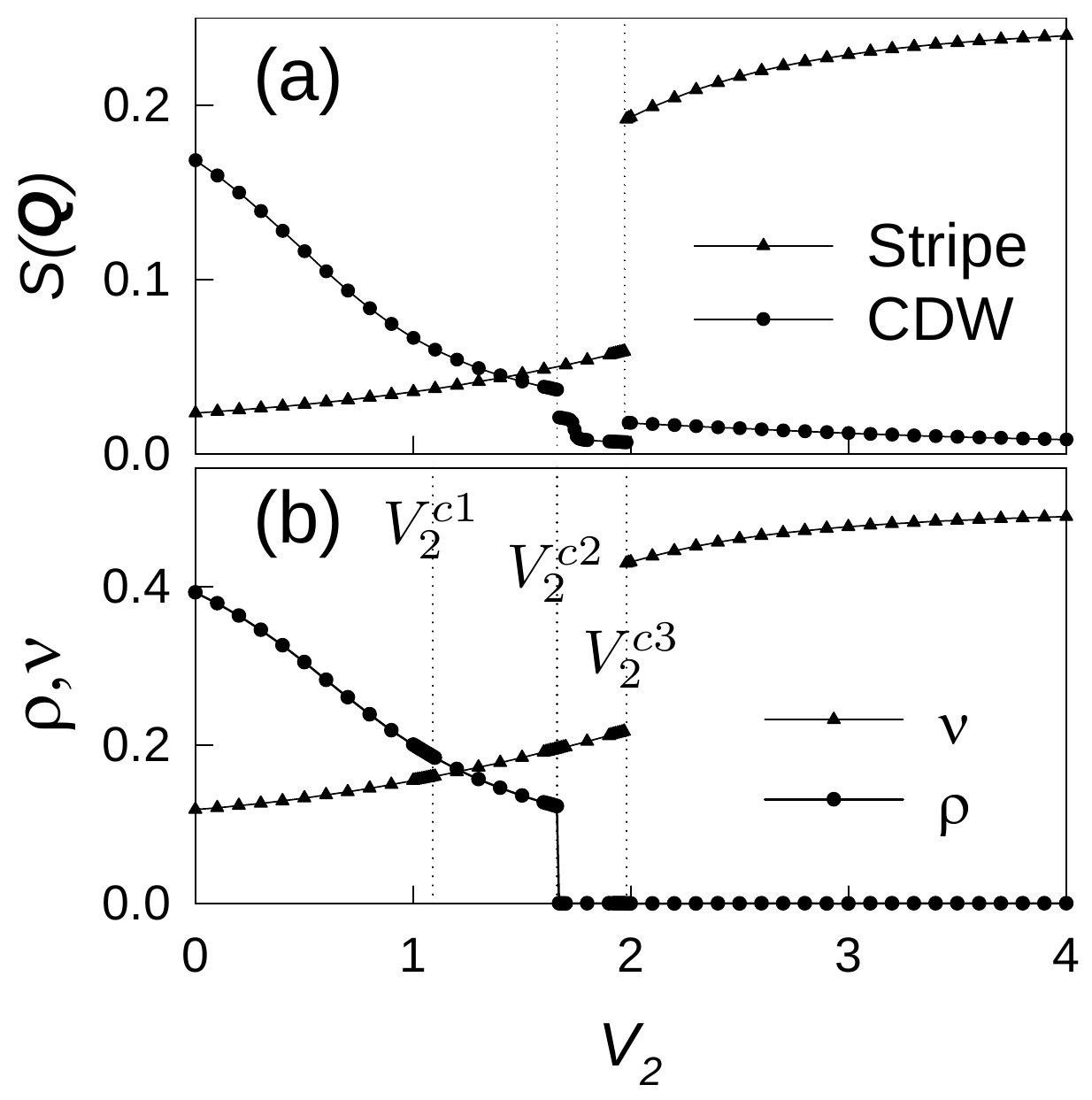}
\caption{(a) The static structure factor $S({\bf Q})$ with ${\bf Q}=(\pi,0)$ or ${\bf Q}=(0,\pi)$ for the stripe phase and ${\bf Q}=(\pi,\pi)$ for the CDW phase. (b) The order parameters $\rho, \nu$ vs $V_2$. Here $V_1=2$. The critical points precisely match those in Fig.\ref{en}. }\label{rho}
\end{figure}

Due to the exchange decoupling channel of NNN interaction in the mean-field approximation,  it is suggested that a Dirac semi-metal (SM) can be driven into the quantum anomalous Hall (QAH) phase in the presence of longer-ranged interaction, realizing TMI. Previous results are obtained using mean-field theory, so it is necessary to study this problem using exact numerical methods to give further confirmation. In the following using ED method, we study whether the topological phase can be generated by interactions in a system with gapless Dirac point.

We drop the $t_2$ term and firstly consider the effect of NN and NNN interactions. The total Hamiltonian under study is,
\begin{equation}\label{eq3}
    H_{Tol}=\sum_{\langle ij \rangle}t_1 e^{i\chi_{ij}}c_i^{\dagger}c_j+\sum_{\langle ij \rangle}V_{1}n_{i}n_{j}+\sum_{\langle \langle ij \rangle \rangle}V_{2}n_{i}n_{j}.
\end{equation}
We calculate the Chern number in the $(V_1,V_2)$ plane and the values are all zero, showing that no topological phase is induced when only the NN and NNN interactions are included.

Then we calculate the six lowest states and study the possible phases in the system. Figure \ref{en} shows a representative energy spectrum at fixed $V_1=2$. For small $V_2$ the ground-state is nearly two-fold degenerate and gapped from other states, denoting the states as $\psi_1$ and $\psi_2$. Two new states can be constructed using a linear combination: $\psi' _1=\frac{1}{\sqrt{2}}(\psi_1+\psi_2)$, $\psi' _2=\frac{1}{\sqrt{2}}(\psi_1-\psi_2)$, which correspond to the two configurations with staggered CDW order (Fig.\ref{lattice}). The other four eigenstates have two close eigenvalues and each is two-fold degenerate. Also by a proper linear combinations,  four new states can be constructed and each corresponds to one of the four configurations with the stripe order. As $V_2$ increases the gap decreases and vanishes at a critical value $V_2^{c1}$, after which the ground state becomes the one with the stripe order [see Fig.\ref{en}(b) for details]. Then the states with CDW order have higher energies. From another critical value $V_2^{c2}$ the split between the two CDW states begins to decrease and vanishes soon [see Fig.\ref{en}(c)]. As $V_2$ further increases, the split between the four stripe states begins to decrease at the critical value $V_2^{c3}$. Meanwhile the gap between the CDW and stripe states vanishes. After that the gap reopens and there is no split any more. The ground-state is exactly four-fold degenerate and with the stripe order.

\begin{figure}[htbp]
\centering
\includegraphics[width=7cm]{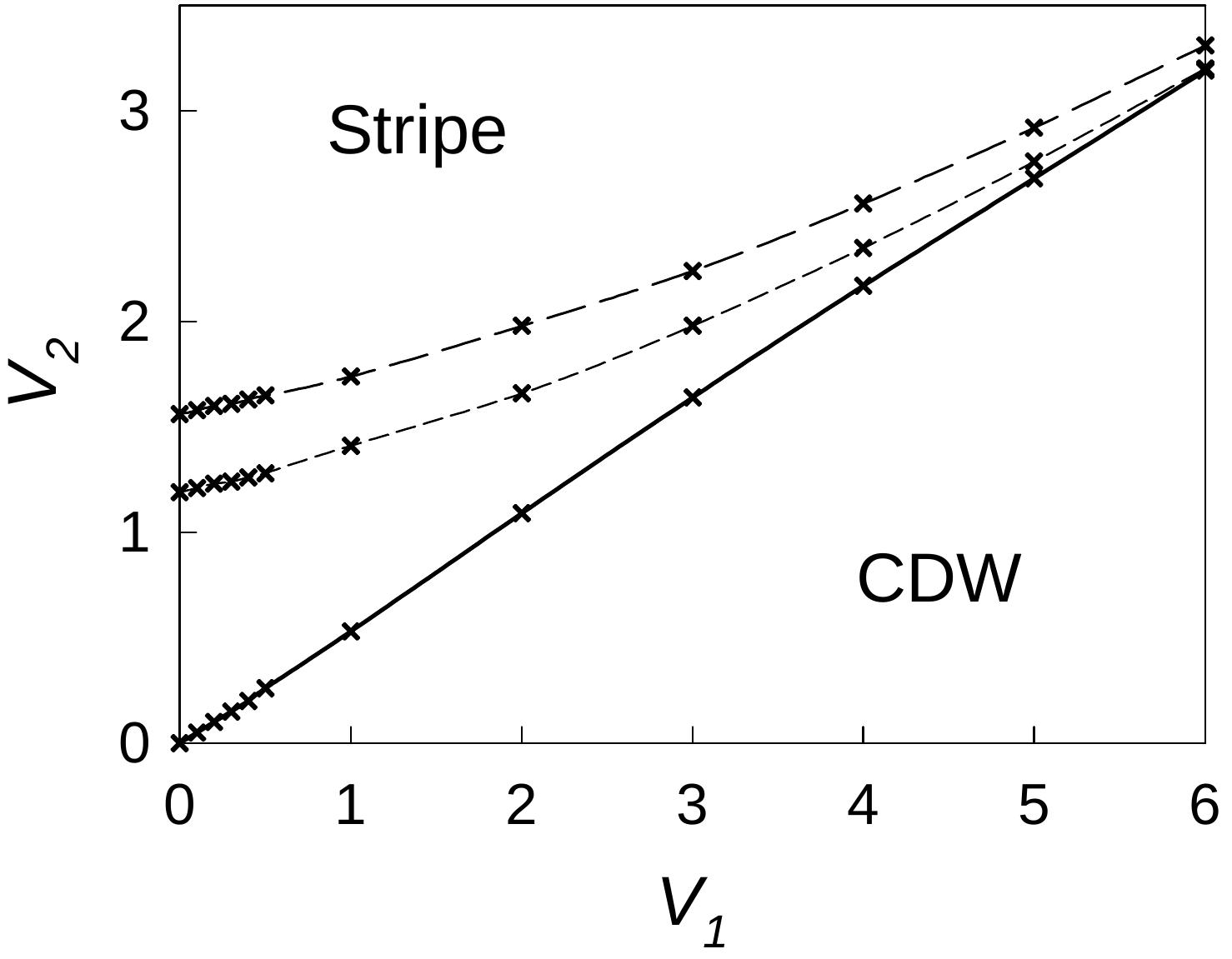}
\caption{The phase diagram in $(V_1,V_2)$ plane obtained from ED calculations. The black solid line is determined by $V_2^{c1}$. The short-dashed line is determined by $V_2^{c2}$. The long-dashed line is determined by $V_2^{c3}$.}\label{phase1}
\end{figure}

For the six lowest states, two of them are with CDW order and four are with stripe order. To characterize different orders, we study SSF:
\begin{equation*}
    S({\bf Q})=\frac{1}{N^2}\sum_{j,k}e^{-i{\bf Q}\cdot ({\bf r}_j-{\bf r}_k)}\langle n_j n_k\rangle.
\end{equation*}
which measure the CDW order when ${\bf Q}=(\pi,\pi)$ while the stripe order when ${\bf Q}=(\pi,0)$ or $(0,\pi)$. At fixed $V_1=2$, $S({\bf Q})$ vs $V_2$ is shown in Fig.\ref{rho}(a). The SSF for the corresponding orders has finite values and shows discontinuities precisely at the critical $V_2$ marked in Fig.\ref{en}. In these ordered states, the average density on each site can be expressed as $\langle n_i\rangle=1/2+\rho (-1)^{i_x+i_y}$ (CDW order) or $\langle n_i\rangle=1/2+\nu (-1)^{i_x(i_y)}$ (stripe order). In Fig.\ref{rho}(b) we plot $\rho,\nu$ as a function of $V_2$ and they also show discontinuous at the critical points $V_2^{c2}, V_2^{c3}$. At $V_2^{c2}$, $\rho$ jumps to zero \cite{note2} while at $V_2^{c3}$, $\nu$ jumps to nearly $1/2$. Thus in combination with the SSF, the nature of the ground states is clearly identified. We scan in the $(V_1,V_2)$ plane and give the critical lines determined by $V_2^{c1}, V_2^{c2}, V_2^{c3}$, which is shown in Fig.\ref{phase1}.

Then we can conclude the phase diagram when the NN and NNN interactions are included in the massless Dirac fermion. Basically the ground state is topological trivial with CDW or stripe order, which is separated by the solid line in Fig.\ref{phase1}. In the upper region of the long-dashed line, the ground state is with the stripe order and are four-fold degenerate. Between the solid and long-dashed lines, the ground state are also with the stripe order, however the eigen-energies of the four stripe states are split to two values, each of which is two-fold degenerate. Compared to some other works \cite{tmi4,neg1,neg2}, the predicted SM phase is not identified. At $V_1=V_2=0$ the system is SM with two Dirac points, each of which is two-fold degenerate. In the ED calculation, the phase is identified by a six-fold degenerate ground-state at half-filling. However when even a very small NN or NNN interaction is introduced, the six-fold degeneracy is split and the ground-state becomes gapped with CDW or stripe order.

\begin{figure}[htbp]
\centering
\includegraphics[width=7cm]{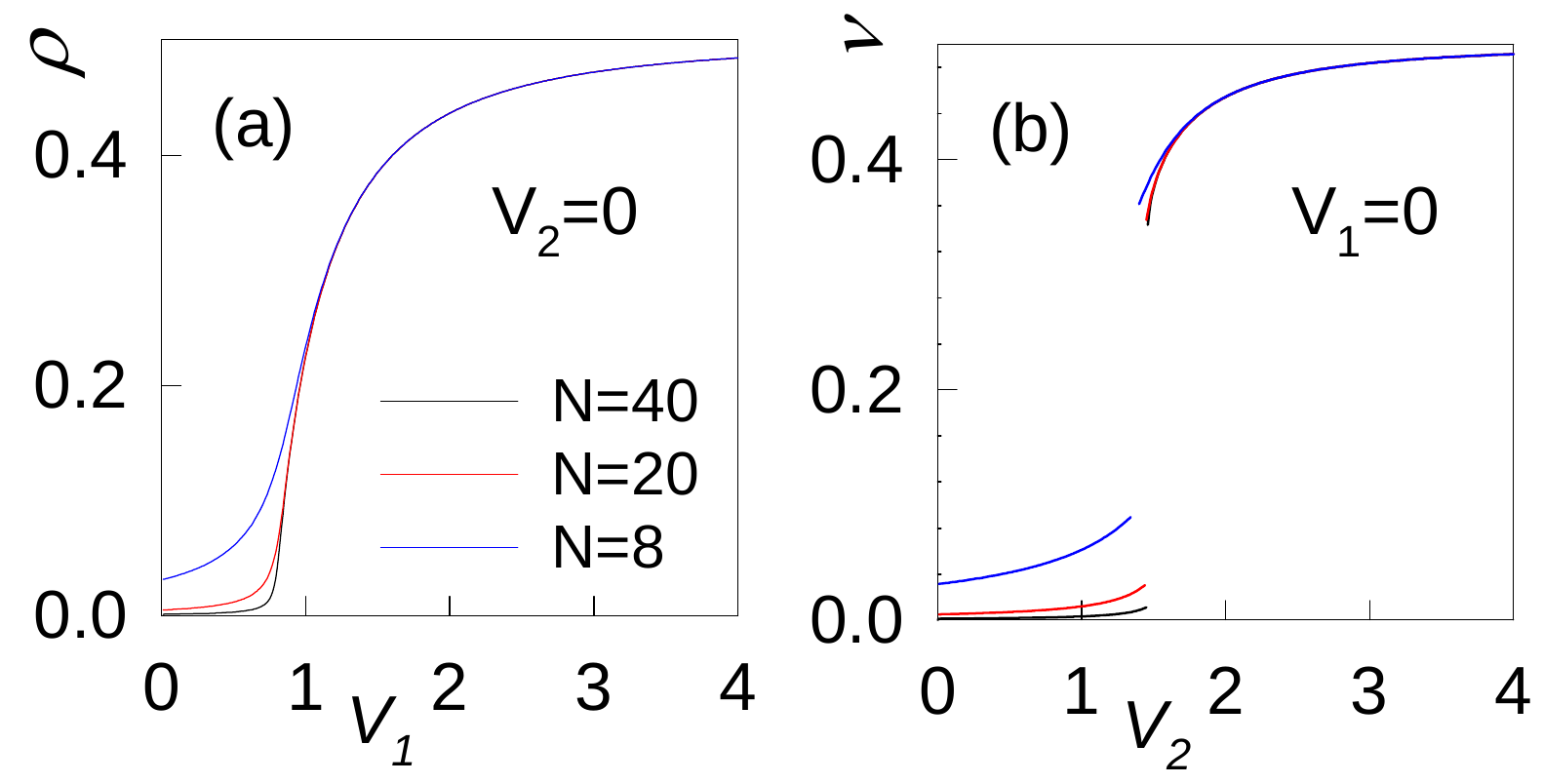}
\caption{(Color on line) The self-consistent order parameters: (a) $\rho$; (b) $\nu$, in the mean-field approximation.} \label{fig8}
\end{figure}

The absence of the SM phase may be due to the finite-size effect. As shown in Fig.\ref{fig8}, in the mean-field approximation, although the self-consistent order parameters have finite values on small sizes for weak interactions, they tend to be zero as the size is increased. Also the mean-field results shown that the transition from the SM to the CDW phase is continuous, while the transition to the stripe phase is not. This implies that the discontinuity at $V_2^{c3}$ in Fig.\ref{rho} may be the boundary between the SM and stripe phases. However due to continuous transition between the SM and CDW phase, it can not be identified by our ED calculations.

We want to emphasize that several recent studies of the effect of on-site Hubbard interaction in the honeycomb Dirac fermions conclude differently on the SM phase \cite{neg1,neg2,lu,ctra1}. The contradictions contain: 1, the SM phase exists below a critical interaction or only at the non-interacting case; 2, if the SM phase exists below a critical interaction, is the Fermi velocity renormalized, or not? Our above results provide some clues on the stability of the Dirac points to the weak interactions. Though the finite-size effect makes the multi-degenerate SM phase splitting into symmetry breaking phases and can not be identified directly, the splitting of the symmetry breaking phases may suggest the existence of the SM phase indirectly. This problem is still worth of further exploration using large-scale numerical methods.

Till now we discuss the phase diagram in the presence of NN and NNN interactions and no topological phase is found. Next we add NNNN interaction $H_{NNNN}=\sum_{\langle \langle \langle ij \rangle \rangle \rangle}V_{3}n_{i}n_{j}$ to Eq.(\ref{eq3}).
We calculate the Chern number in $(V_1,V_2)$ plane at different fixed $V_3$ and don't find any region showing topological phase.

\section{Conclusion and discussion}
In conclusion, we study the interaction effect in two-dimensional Dirac fermions of the $\pi-$ flux model. Firstly we consider the interaction effect in the topological phase. Using ED method we calculate the Chern number, the energy spectrum, the fidelity metric, the SSF and the electrons' distribution of the ground-state. From these calcualtions, we obtain the phase diagram in the $(V_1,V_2)$ plane and find the topological phase persist in a finite region of the phase diagram. The results show that the topological phase is robust to small interactions. To understand the phase diagram, we also present a qualitative explanation using the mean-field approximation. Next we study the possibility of dynamically generate topological phase in massless Dirac fermions. When only $V_1$ and $V_2$ are present, the system may be in the CDW or stripe phase, which are all topological trivial. When the NNNN interaction is further included, the topological phase is still absent in our calculations.

Specially one of our main results is the absence of the topological phase predicted by mean-field approximation. The deviations may be attributed to the small size on which the ED simulations are performed. However it should be noted that previous and the present works have shown that even for small sizes the topological property already can manifest itself well \cite{kmu3}. So our results can demonstrate that at least on small sizes the topological phase does not be generated following the mechanism of the mean-field theory. Considering the strong quantum fluctuations in two dimensions which may make the mean-field approximation break down, large-scale numerical simulations are expected to give further clarification on this important problem.

It has been proposed that the Hamiltonian Eq.(\ref{eq1}) can be realized in two-dimensional electron gas sandwiched between two type-II superconducting films \cite{model1}. The interaction in such device is weak, but it can provide a realistic chance to study the weak-coupling regime. Another possible experimental platform is the optical lattice, where the geometry and parameters of the model can be tunable. There have been intriguing theoretical proposals of creating effective magnetic fields in square optical lattices \cite{ol1}. Moreover a staggered magnetic field has been realized experimentally recently \cite{ol2}. It is hopeful that large uniform magnetic flux is created. So in the near future this problem may be clarified in these promising experiments.

\section{Acknowledgements}
HG is supported by NSFC under Grant Nos. 11274032, 11104189, FOK YING TUNG EDUCATION FOUNDATION and Program for NCET. ZC is supported by  NSFC under Grant No. 11274033. SS is supported by the Research Grant Council of Hong Kong under Grant No. N HKU748/10. SP is supported by NSFC under Grand Nos. 11074023 and 11274044, and funds from the Ministry of Science and Technology of China under Grand Nos. 2011CB921700 and 2012CB821403.


\begin{thebibliography}{10}
\bibitem[*] {add} Email address: hmguo@buaa.edu.cn.
\bibitem{rev1}
J.E.~Moore, Nature {\bf 464}, 194 (2010).
\bibitem{rev2}
M.Z.~Hasan, C.L.~Kane, \rmp, {\bf 82}, 3045 (2010).
\bibitem{rev3}
Xiao-Liang Qi and Shou-Cheng Zhang, \rmp, {\bf 83}, 1057 (2011).
\bibitem{rev4}
X.-L. Qi and S.-C. Zhang, Phys. Today {\bf 63}, 33 (2010).

\bibitem{int1}
M. Hohenadler, F. F. Assaad, J. Phys.: Condens. Matter {\bf 25}, 143201 (2013).
\bibitem{int2} Dmytro Pesin and Leon Balents, Nature Physics {\bf 6}, 376 (2010).

\bibitem{km1} C.L. Kane and E.J. Mele, \prl {\bf 95}, 146802 (2005).
\bibitem{km2} C.L. Kane and E.J. Mele, \prl {\bf 95}, 226801 (2005).

\bibitem{kmu1} M. Hohenadler, T. C. Lang, and F. F. Assaad, \prl {\bf 106}, 100403 (2011); M. Hohenadler, Z. Y. Meng, T. C. Lang, S. Wessel, A. Muramatsu, and F. F. Assaad, \prb {\bf 85}, 115132 (2012).
\bibitem{kmu3} C. N. Varney, K. Sun, M. Rigol, and V. Galitski, \prb 82, 115125 (2010); C. N. Varney, Kai Sun, Marcos Rigol, and Victor Galitski, \prb {\bf 84}, 241105 (2011).
\bibitem{kmu5} Dong Zheng, Guang-Ming Zhang, and Congjun Wu, \prb {\bf 84}, 205121 (2011).
\bibitem{kmu6} Youhei Yamaji and Masatoshi Imada, \prb {\bf 83}, 205122 (2011).
\bibitem{kmu7} Shun-Li Yu, X. C. Xie, and Jian-Xin Li, \prl 107, 010401 (2011).
\bibitem{kmu8} Wei Wu, Stephan Rachel, Wu-Ming Liu, and Karyn Le Hur, \prb {\bf 85}, 205102 (2012).


\bibitem{th1} Jun Wen, Mehdi Kargarian, Abolhassan Vaezi, and Gregory A. Fiete, \prb {\bf 84}, 235149 (2011).
\bibitem{th2} Stephan Rachel and Karyn Le Hur, \prb {\bf 82}, 075106 (2010).
\bibitem{th3} Dung-Hai Lee, \prl {\bf 107}, 166806 (2011).
\bibitem{th4} Christian Griset and Cenke Xu, \prb {\bf 85}, 045123 (2012).

\bibitem{tmi1} S. Raghu, Xiao-Liang Qi, C. Honerkamp, and Shou-Cheng Zhang, \prl {\bf 100}, 156401 (2008).
\bibitem{tmi2} Yi Zhang, Ying Ran, and Ashvin Vishwanath, \prb {\bf 79}, 245331 (2009).
\bibitem{tmi3} Jun Wen, Andreas R¨¹egg, C.-C. Joseph Wang, and Gregory A. Fiete, \prb {\bf 82}, 075125 (2010).
\bibitem{tmi4} C. Weeks and M. Franz, \prb {\bf 81}, 085105 (2010).

\bibitem{tmi6} Wei-Feng Tsai, Chen Fang, Hong Yao, JiangPing Hu, arXiv:1112.5789 (2011).
\bibitem{tmi7} Qin Liu, Hong Yao, Tianxing Ma, \prb, {\bf 82}, 045102 (2010).
\bibitem{tmi8} A. Dauphin, M. M¨¹ller, and M. A. Martin-Delgado, \pra, {\bf 86}, 053618 (2012).
\bibitem{tmi9} Moyuru Kurita, Youhei Yamaji, and Masatoshi Imada, J. Phys. Soc. Jpn. {\bf 80}, 044708 (2011).
\bibitem{tmi10} Kai Sun, Hong Yao, Eduardo Fradkin, and Steven A. Kivelson, \prl {\bf 103}, 046811 (2009).


\bibitem{model1} G. Rosenberg, B. Seradjeh, C. Weeks, and M. Franz, \prb {\bf 79}, 205102 (2009).
\bibitem{model2} Huai-Ming Guo and Shi-Ping Feng, Chinese Physics B {\bf 21}, 077303 (2012).

\bibitem{jiexi} Shun-Qing Shen, Topological Insulators, (Springer, 2012).

\bibitem{method} T. Fukui, Y. Hatsugai, and H. Suzuki, J. Phys. Soc. Jpn. 74, 1674 (2005).

\bibitem{fu} Liang Fu and C. L. Kane, \prb {\bf 76}, 045302 (2007).
\bibitem{guo1} H.-M. Guo and M. Franz, \prb {\bf 80}, 113102 (2009); H.-M. Guo and M. Franz, \prl {\bf 103}, 206805 (2009).

\bibitem{note1} The knowledge of the parity can determine the $Z_2$ topological invariant of time-reversal invariant band insulators. Here the Hamiltonian Eq.(\ref{eq1}) breaks time-reversal symmetry, but the parity is still related to the topological properties of the system. For systems with spin $S_z$ conservation, nontrivial $Z_2$ also means that each spin copy has a nonzero Chern number. Since the parity of each spin copy is identical to that of the whole system, the parity is related to the Chern number of time-reversal breaking two-dimensional systems with inversion symmetry.

\bibitem{note2} When $\rho=0$, the SSF of CDW order has a small value, which is due to the fact that $\langle n_j n_k\rangle\neq \langle n_j \rangle\langle n_k\rangle$ since the number operator on a sole site $n_j$ does not communicate with the Hamiltonian.

\bibitem{neg1} Z. Y. Meng, T. C. Lang, S. Wessel, F. F. Assaad, and A. Muramatsu, Nature {\bf 464}, 847 (2010).

\bibitem{neg2} Igor F. Herbut, Vladimir Juricic, and Bitan Roy, \prb {\bf 79}, 085116 (2009).

\bibitem{lu} Rong-Qiang He and Zhong-Yi Lu, \prb {\bf 86}, 045105 (2012).
\bibitem{ctra1} Wei Wu, Yao-Hua Chen, Hong-Shuai Tao, Ning-Hua Tong, and Wu-Ming Liu, \prb {\bf 82}, 245102 (2010).

\bibitem{ol1} D. Jaksch and P. Zoller, New J. Phys. {\bf 5}, 56 (2003).
\bibitem{ol2} M. Aidelsburger, M. Atala, S. Nascimb¨¨ne, S. Trotzky, Y.-A. Chen, and I. Bloch, \prl {\bf 107}, 255301 (2011).















\end{thebibliography}
\end{document}